\documentclass[%
 reprint,
showpacs,
 amsmath,amssymb,
 aps,
prb,
]{revtex4-1}

\usepackage{graphicx}
\usepackage{dcolumn}
\usepackage{bm}

\begin{document}

\preprint{APS/123-QED}

\title{Griffiths phase in solid-solution of ferromagnetic manganite and cobaltite}

\author{D. Bhoi$^1$, N. Khan$^1$, A.Midya$^1$, A. Hassen$^{1,2}$ , P.Mandal$^1$, P.Chudhury$^3$}
\affiliation{$^1$Saha Institute of Nuclear Physics, 1/AF Bidhannagar, Calcutta 700 064, India\\
            $^2$Physics Department, Faculty of Science, Fayoum University, 63514 El Fayoum, Egypt\\
            $^3$Central Glass and Ceramic Research Institute, 196 Raja S. C. Mullick Road, Calcutta 700 032, India}

\date{\today}

\begin{abstract}
We report the existence of Griffiths phase (GP) over a wide range of $x$ in La$_{0.6}$Sr$_{0.4}$Mn$_{1-x}$Co$_x$O$_3$, the solid solution of ferromagnetic (FM) La$_{0.6}$Sr$_{0.4}$MnO$_3$ and La$_{0.6}$Sr$_{0.4}$CoO$_3$, from  magnetization  measurements. In this compound, GP arises due to the quenching of randomly distributed Co-O-Mn antiferromagnetic bonds in the  FM background. In contrary to divalent doped manganites, GP in the present system can exist entirely in the metallic  state above $T_C$ (for $x$$<$0.10). Based on the present study, a magnetoelectronic phase diagram is drawn.
\end{abstract}

\maketitle

The undoped parent  compounds of rare-earth transition metal oxides  are magnetoelectronically homogeneous because of the single valence state of the transition metal ion with antiferromagnetic (AFM) interaction. With doping, these systems become intrinsically inhomogeneous due to the random distribution of cations of different sizes and valence/spin states and the strong competition between different ordering tendencies.  Accumulated experimental data from various high resolution probes confirm the electronic and magnetic inhomogeneities owing to the phase separation in microscopic length scale \cite{dago,tere,he}. Among the various forms of   phase separation, the presence of "preformed" ferromagnetic (FM)  clusters  well above the  Curie temperature ($T_C$) seems to be of particular importance \cite{dago,tere,he}. Preformed clustering also emerges from theoretical descriptions.  Within the context of quenched disorder scenarios, the emergence  of a "Griffiths-like" clustered phase  has been predicted  where the coexistence of two competing ordered phases enhances the formation of this phase  below a characteristic temperature $T_{GP}$ \cite{burgy}.  Below $T_{GP}$ and above $T_C$, the quenched disorder system is in between the completely disordered paramagnetic (PM) and FM ordered states. To describe this phase,  Griffiths considered  a percolation like problem in which a fraction ($p$) of the lattice sites  is occupied by magnetic ions with a nearest-neighbor FM interaction of strength $J$ and the rest ($1$$-$$p$) is occupied by nonmagnetic ions with strength 0 \cite{griff}. Above the percolation threshold ($p_c$), FM order sets in at $T_C$($p$) which is below the ordering temperature of the undiluted system $T_C$($p$$=$1), and in the region $T_C$($p$)$<$$T$$<$$T_C$($p$$=$1), the thermodynamic properties  are nonanalytical due to the formation of  short-range ordered clusters. $T_C$($p$$=$1)  is, therefore, the temperature below which this "Griffiths phase (GP)" forms and coined  to $T_{GP}$ \cite{bray}. \\

In spite of theoretical prediction about four decades ago and its subsequent developments \cite{griff,bray}, GP has not been realized experimentally until recently  in heavy fermions, manganites, layered cobaltites and intermetallics  \cite{cast,sala,rama,deisen,jiang,pram,shim,mag}. So far, in manganites, GP is observed to appear only in the insulating state.  The occurrence of GP in the insulating  state is not  a  mere coincidence. It has been argued from the theoretical ground  and  a wealth of experimental data that the MI transition  in manganites appears in tandem with GP, both being consequences of the same percolation effect \cite{sala,kri}. In contrary to  manganites, there is no experimental evidence on the formation of GP  in the canonical double-exchange (DE) system La$_{1-x}$Sr$_{x}$CoO$_3$, though, the existence of FM clusters  and percolative ferromagnetism are more common in cobaltites  \cite{he}. He {\it et al}. extensively studied the Griffiths phase aspect in La$_{1-x}$Sr$_{x}$CoO$_3$ and observed that FM clusters are formed well above  $T_C$ as in the case of manganites but they are  non-Griffiths-like in nature. This  raises a question that, although the Griffiths model may apply to many systems with quenched disorder, it is not   applicable to all randomly doped transition metal oxides, particularly to three dimensional cobaltites \cite{he}. In this work, we demonstrate the existence of GP in La$_{0.6}$Sr$_{0.4}$Mn$_{1-x}$Co$_x$O$_3$ (LSMCO) over a wide range of $x$ from magnetic measurements. We observe that GP appears and exists entirely in the metallic state (d$\rho$/d$T$$>$0) for $x$$<$0.10 and $x$$>$0.85. To the best of our knowledge, this is the first experimental evidence on the existence of GP   in the metallic state in manganite and also in perovskite cobaltite La$_{0.6}$Sr$_{0.4}$CoO$_3$. \\

Samples of nominal compositions La$_{0.6}$Sr$_{0.4}$Mn$_{1-x}$Co$_x$O$_3$ (0$\leq$$x$$\leq$1.0) were prepared by the standard solid-state reaction method.  The Rietveld x-ray analysis reveals that these materials are single phase and the diffraction patterns can be indexed with rhombohedral unit cell for all compositions. Energy dispersive x-ray confirms that they are chemically homogeneous with composition close to the nominal ones. Magnetic measurements were done using a SQUID magnetometer (Quantum Design).  The magnetotransport measurements were performed  by  standard dc four-probe method down to 1.5 K  with  fields up to 5 T.\\

\begin{figure}
  \includegraphics[width=0.4\textwidth]{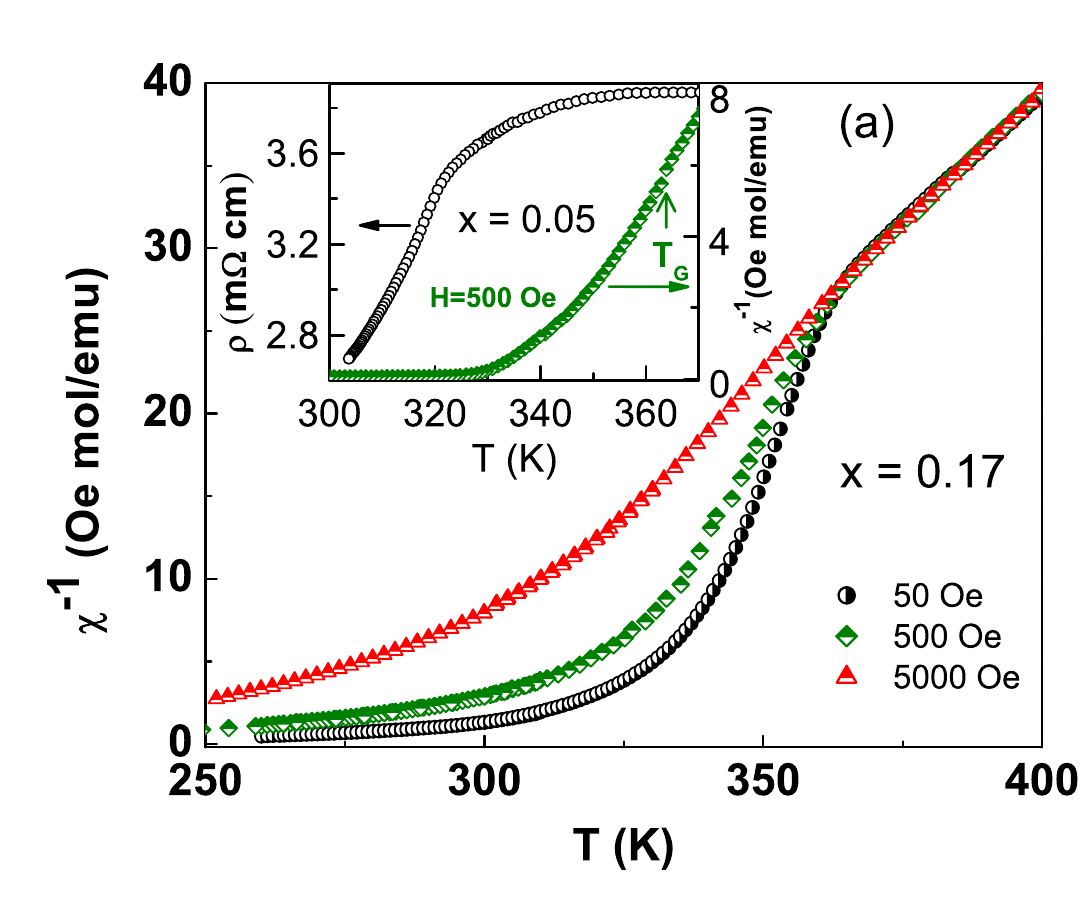}
  \includegraphics[width=0.4\textwidth]{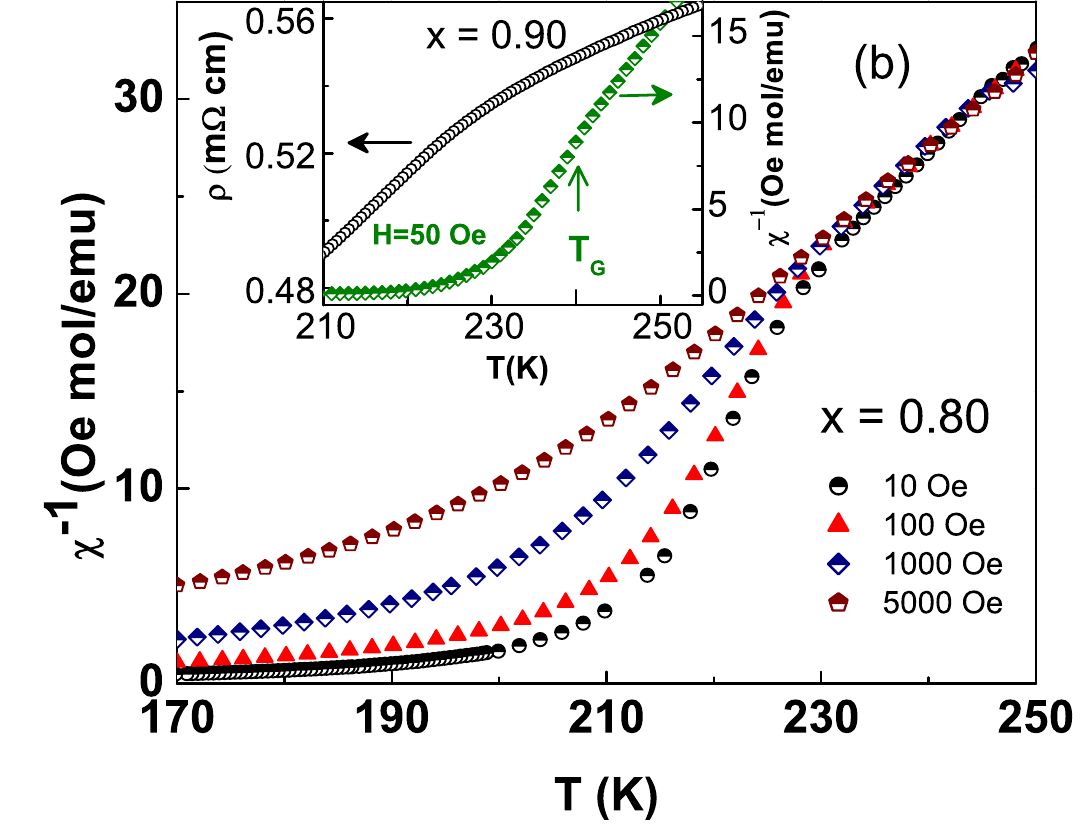}
  \includegraphics[width=0.4\textwidth]{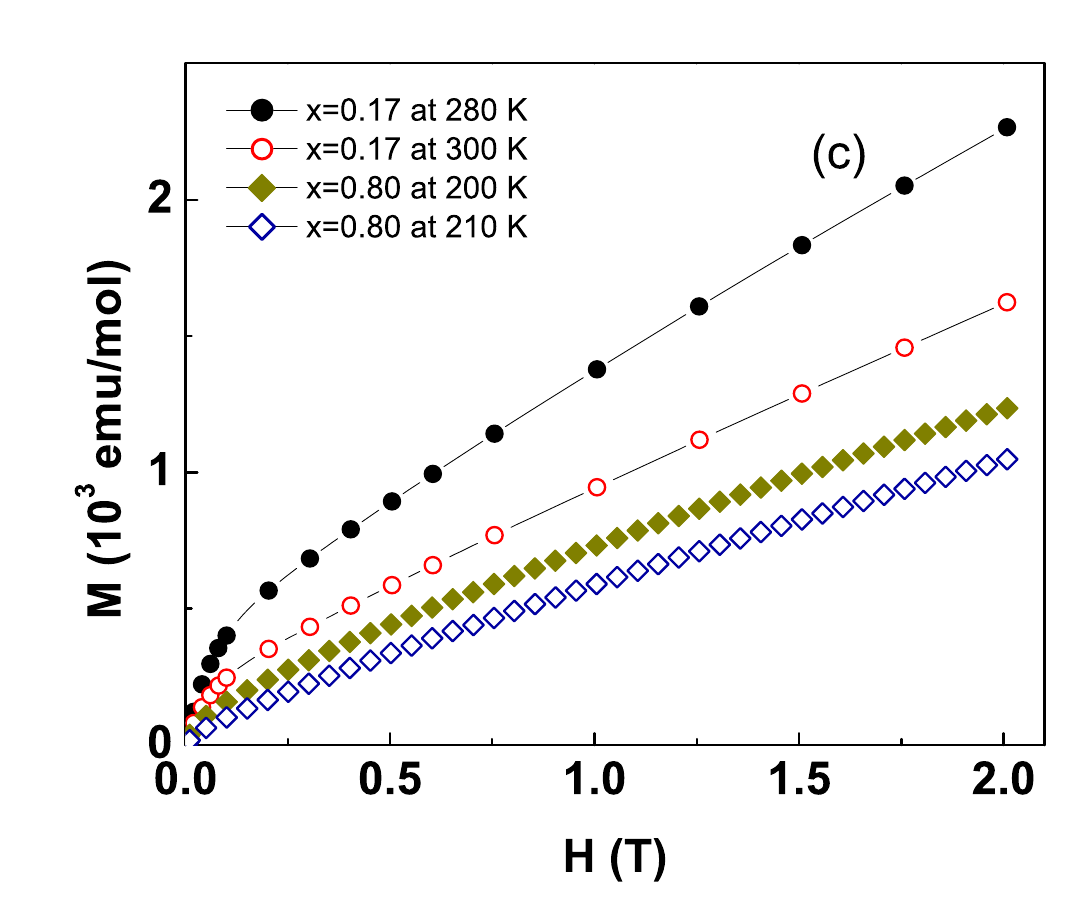}
  \caption{Temperature dependence of inverse susceptibility ($\chi^{-1}$) at different fields for La$_{0.6}$Sr$_{0.4}$Mn$_{1-x}$Co$_x$O$_3$ with $x$$=$ 0.17 (a) and 0.80 (b). Insets show $\chi^{-1}$($T$) and $\rho$($T$) plots for $x$$=$0.05 and 0.90 samples. $M$($H$) plot for $x$$=$0.17 and 0.80 samples at some selected temperatures for $T_C$$<$$T$$<$$T_G$ (c).}\label{Fig.1}
\end{figure}

In perovskite manganites, the presence of a Griffiths-like clustered phase  within the PM matrix  has been demonstrated from  the macroscopic magnetization measurements \cite{sala}. In Fig. 1, we  have shown the inverse dc susceptibility $\chi^{-1}$ ($=$$H/M$) vs $T$ plots for LSMCO for  compositions $x$$=$0.05, 0.17, 0.80 and 0.90 as representatives. It is clear from the figure that $\chi^{-1}$($T$) does not follow the usual Curie-Weiss (CW) law above $T_C$. As the temperature decreases, $\chi^{-1}$ displays a downward deviation from the CW law  below  a characteristic temperature $T_{GP}$ and a positive curvature in the region $T_C$$<$$T$$<$$T_{GP}$. The faster decrease of $\chi^{-1}$ below $T_{GP}$ signals the onset of short-range FM correlation well above $T_C$ and this is considered to be a hallmark of Griffiths singularity \cite{sala,rama,deisen,jiang,pram,shim,mag}. For further confirmation that the observed phenomenon is actually due to the Griffiths singularity, we have measured $\chi$ for different $H$. $\chi$ is expected to increase  over the CW behavior with decreasing field strength, at least for low $H$, where the susceptibility of the clusters is dominant. This behavior is clearly reflected (Fig. 1). At high fields, the contribution from  PM matrix is significant, as a result, $\chi^{-1}$($T$) becomes almost linear in $T$ above $T_C$. The downward deviation in $\chi^{-1}$($T$) from the CW law  has been observed in several manganites and this behavior has been attributed  to the formation of  Griffiths phase associated with FM clusters  \cite{sala,rama,deisen,jiang,pram}. Formation of nanosize FM clusters in Tb$_5$Si$_2$Ge$_2$  has been demonstrated by small angle neutron scattering (SANS) in the region $T_C$$<$$T$$<$$T_{GP}$ \cite{mag}. The different plateaux in  $\chi^{-1}$($T$) curve of Tb$_5$Si$_2$Ge$_2$ below $T_{GP}$ have been correlated to FM clusters of different sizes. From the slope of  $\chi^{-1}$($T$) for the present sample, we can deduce the value of $\mu_{eff}$   to get a rough estimation on cluster size as it was done in the case of other manganites \cite{borg}. Above $T_{GP}$,  the calculated value of $\mu_{eff}$ is close to the average spin only moment. However, except in a narrow range of $T$ below $T_{GP}$, we observe that $\mu_{eff}$ increases rapidly with decreasing temperature  for $T$$<$$T_{GP}$ and its value is much larger than the theoretical one.  For example, $\mu_{eff}$$\sim$25 $\mu_{B}$ at $T$$=$280 K and $H$$=$50 Oe for the $x$$=$0.17 sample which corresponds to $S$$\sim$12.  This is an indication of the formation of  cluster of few magnetic ions together with a FM correlation. As in the case of Tb$_5$Si$_2$Ge$_2$, we observe that the cluster size diverges as $T$ approaches $T_C$.\\

\begin{figure}
  \includegraphics[width=0.45\textwidth]{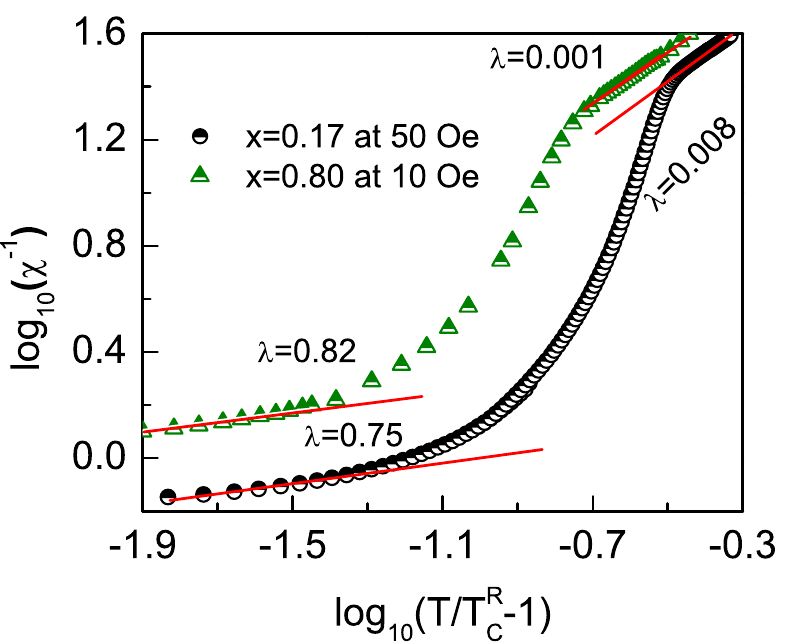}\\
  \caption{(Color online) log$_{10}$$\chi^{-1}$ vs log$_{10}$($T/T_C^R$-1) plots for La$_{0.6}$Sr$_{0.4}$Mn$_{1-x}$Co$_x$O$_3$ with  $x$$=$ 0.17 and 0.80.  The solid lines are fit to data using Eq. (1) in the text.}\label{Fig.2}
\end{figure}

Reported  data on  La$_{0.6}$Sr$_{0.4}$MnO$_3$  and La$_{0.6}$Sr$_{0.4}$CoO$_3$ compounds  show that $\chi^{-1}$($T$) obeys the CW law well above $T_C$ but a positive curvature  appears as $T$ approaches  $T_C$ \cite{he,souza}. However, there is a qualitative difference in the nature of $\chi$($T$) for these compounds and that for LSMCO.  $\chi$ is observed to increase slower than the CW law with decreasing $T$ for both La$_{0.6}$Sr$_{0.4}$MnO$_3$ and La$_{0.6}$Sr$_{0.4}$CoO$_3$ while in the case of LSMCO it is just opposite. Besides this, unlike LSMCO,  $\chi$ for La$_{0.6}$Sr$_{0.4}$CoO$_3$ above $T_C$ is independent of $H$, though, the formation of FM clusters in this compound  at a temperature as high as 360 K has been reported using different experimental techniques \cite{he}. Another important characteristic of GP  is that the system as a whole does not exhibit any spontaneous magnetization because these clusters are finite in size. To verify this,  $M$($H$) has been measured in the region $T_C$$<$$T$$<$$T_{GP}$. As expected, $M$ increases rapidly with $H$ at low fields as in the case of a ferromagnet but at high fields where the PM matrix dominates $M$, the increase is approximately linear in $H$ [Fig. 1(c)]. The Arrott plots of $M$($H$) data (not shown) demonstrate that nonzero spontaneous magnetization exists only below  $T_C$. One can see that the contribution from FM clusters to $M$ does not get masked by  the PM matrix even up to a field of few kOe; indicating the stability of GP.   \\

The susceptibility in the Griffiths phase is characterized by a power law \cite{cast,sala},
\begin{equation}
\chi^{-1}(T)=A(T-T_C^R)^{1-\lambda},
\end{equation}
where 0$\leq$$\lambda$$<$1 and $T_C^R$ is the critical temperature of the random ferromagnet. To calculate $\lambda$, we have plotted $\chi^{-1}$ vs  ($T/T_C^R-1$) on log-log scale. For the accurate determination of $\lambda$, $T_C^R$  has been adjusted in such a way that $\chi^{-1}$($T$) above $T_{GP}$ be linear for $\lambda$ close to zero, i.e., the usual CW behavior is recovered in the PM state. Figs. 2(a) and (b) show such plots for two representative compositions, viz, $x$$=$0.17 and 0.80, wherein the fitted values of $\lambda$ are 0.75 and 0.80, respectively. $\lambda$ for other compositions is also close to these values. Thus, the value of $\lambda$ for the present system is close to those reported for other manganites \cite{sala,jiang,pram}.  According to the theoretical model, above $p_c$, $\chi$ should show a power-law divergence as $T$ approaches $T_C$ \cite{cast,sala}.  However, the analysis of experimental data on several systems using Eq. (1) reveals that $\chi$ diverges at a temperature $T_C^R$, which is higher than $T_C$ \cite{sala,pram}. Though, $T_C$ and $T_C^R$ in LSMCO are close to each other for small and large values of $x$, the difference ($T_C^R$$-$$T_C$) increases with the increase of doping concentration. In the Mn-rich side, this difference becomes as high as 100 K at $x$=0.25. The difference between $T_C$ and $T_C^R$ in the Co-rich side  is  not as significant as in the Mn-rich side.\\

The  magnetoelectronic phase diagram for the solid solution La$_{0.6}$Sr$_{0.4}$Mn$_{1-x}$Co$_x$O$_3$ is shown in figure 3. One can notice that the phase diagram  is very rich and complex. Several new phases emerge with doping which are not present in either of the undoped compounds, La$_{0.6}$Sr$_{0.4}$MnO$_3$  and La$_{0.6}$Sr$_{0.4}$CoO$_3$. For better understanding, one can  view  this phase diagram as Co doping at Mn site in La$_{0.6}$Sr$_{0.4}$MnO$_3$ for $x$$<$0.50  and Mn doping at Co site in La$_{0.6}$Sr$_{0.4}$CoO$_3$ for $x$$>$0.50.  It is clear from the figure that $T_C$ (defined as the temperature where d$M$/d$T$ exhibits a minimum) shows a nonmonotonous dependence on $x$.  An approximate linear decrease in $T_C$ is observed up to  25\% of Co doping in La$_{0.6}$Sr$_{0.4}$MnO$_3$ and  20\% of Mn doping in La$_{0.6}$Sr$_{0.4}$CoO$_3$. Above these concentrations, $T_C$ increases and reaches a maximum at around $x$$=$0.50 where Co and Mn contents are same.  In order to understand the nature of the magnetic ground state, we have meticulously examined the magnetic properties of La$_{0.6}$Sr$_{0.4}$Mn$_{1-x}$Co$_x$O$_3$ and observed that the system is highly inhomogeneous  in the intermediate doping region.  The temperature, field and frequency dependence of  $M$ show that a cluster-glass- or spin-glass-like phase dominates the low-temperature magnetic properties  due to the strong competition between FM and AFM interactions.  Indeed, there exists no well-defined FM-PM transition for compositions 0.30$<$$x$$<$0.70 (i.e., the minimum in d$M$/d$T$ curve is quite shallow). In this context, we would like to mention that the $T_C$($x$) curve for La$_{2/3}$Ba$_{1/3}$Mn$_{1-x}$Co$_x$O$_3$ is also 'W'-like  and a glassy magnetic phase with vanishingly small magnetic moment emerges in the intermediate doping region \cite{troy}. In LSMCO, the existence of GP has been detected in two disjoint regions, one for $x$$\leq$0.40 and the other for $x$$\geq$0.75. From the field dependence of $M$, we observe that the absence of GP in the range 0.40$<$$x$$<$0.75 is due to the small FM volume fraction. In both the regions of GP, $T_{GP}$ is nearly independent of $x$. The intersections of $T_{GP}$($x$) lines with the magnetic ordering boundaries reveal two regular ferromagnets corresponding to $p$$=$1; La$_{0.6}$Sr$_{0.4}$MnO$_3$ with $T_C$$=$364 K  and La$_{0.6}$Sr$_{0.4}$CoO$_3$ with $T_C$$=$240 K. Usually, the phase region bounded by the $T_C$($x$) and $T_{GP}$($x$) lines is known as GP. As mentioned earlier, experimentally  GP  is observed in the temperature interval $T_C^R$$<$$T$$<$$T_{GP}$.  In this sense, the shape of GP in both the regions is triangular, similar to that observed in other perovskite manganites such as La$_{1-x}$Sr$_{x}$MnO$_3$ and La$_{1-x}$Ba$_{x}$MnO$_3$ \cite{deisen,jiang}. \\

\begin{figure}
  \includegraphics[width=0.45\textwidth]{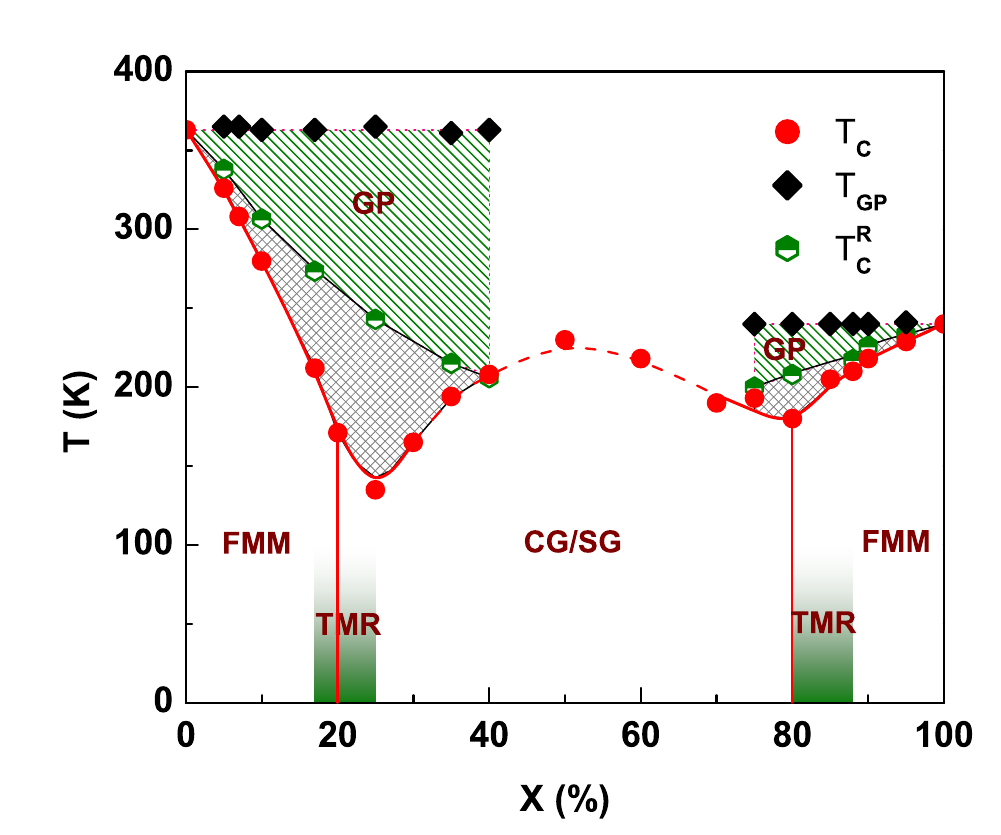}\\
  \caption{(Color online) Magnetoelectronic phase diagram for La$_{0.6}$Sr$_{0.4}$Mn$_{1-x}$Co$_x$O$_3$, wherein the different transition temperatures are plotted against the Co concentration $x$. The doping driven metal to insulator transitions (defined as 1/$\rho$$=$$\sigma$$\rightarrow$0 as $T$$\rightarrow$0) are shown by the vertical red lines at $x$$=$ 0.20 and 0.80. The green-shaded areas bounded by $x$$=$ 0.17 and 0.25 lines and by $x$$=$ 0.80 and 0.88  lines are TMR regions, where  a huge negative magnetoresistance arises due to intergranular tunneling effect at $T$$\ll$$T_C$. The green hatched areas bounded by the $T_{GP}$ and $T_C^R$ lines denote Griffiths phase. Abbreviations used FMM : ferromagnetic metal,  CG : cluster-glass, SG : spin glass, TMR : tunneling magnetoresistance, and GP : Griffiths phase.}\label{Fig.3}
\end{figure}

Mainly, two types of substitutional disorder related to the undersize and oversize effects are important for the occurrence of GP in divalent doped manganites \cite{sala,rama,deisen,jiang,pram}. When La is substituted by the smaller size Ca ion, the local tilting of MnO$_6$ octahedra and the concurrent bending of the active Mn-O-Mn bonds occur in La$_{1-x}$Ca$_x$MnO$_3$. This inhibits the formation of FM metallic bonds associated with DE mechanism  \cite{sala}. On the other hand, the substitution of larger size Ba and Pb for La  produces a local disorder due to the increase of variance  which also inhibits  the formation of FM metallic bonds. In both the cases,  metallic and nonmetallic bonds coexist in the system. The concentration of metallic bond is large at low temperatures $T$$\ll$$T_C$ but it decreases with increasing $T$ \cite{sala}.  As a result,  a FM metal to PM insulator transition occurs when the concentration of metallic bond reaches the percolation threshold at a  temperature close to $T_C$. Phenomenological models based on effective medium approach suggest that both  MI transition and GP are due to the percolative nature of electrical conduction  \cite{sala,kri}. The situation is slightly different for La$_{1-x}$Sr$_x$MnO$_3$ system. Neither the oversize effect nor the undersize effect is significant in La$_{1-x}$Sr$_x$MnO$_3$ and this system does not exhibit any MI transition for 0.30$\leq$$x$$\leq$0.40, as a result,  GP exists only in the static Jahn-Teller (JT) dominated insulating FM regime 0.06$<$$x$$<$0.16 \cite{deisen}. In brief,  GP in manganites occurs in the insulating state only \cite{sala,rama,deisen,jiang,pram}.\\

In order to examine whether the insulating nature of PM state is a prerequisite for the occurrence of GP, we have measured $\rho$($T$) for our samples.  $\rho$ is metallic (d$\rho$/d$T$$>$0) in the PM state for samples with $x$$<$0.10 and $x$$>$0.85 (see insets of Fig. 1). The existence of GP entirely in the metallic  PM state is in stark contrast with previous reports \cite{sala,rama,deisen,jiang,pram} and cannot be explained using the effective medium theory  \cite{sala,kri}.  In the classical Griffiths model, the exchange interactions are distributed randomly, but once distributed, are fixed in the lattice. Obviously, this is not the case for divalent doped manganites. On the other hand, Co(Mn) doping at Mn(Co) site in LSMCO creates Co-O-Mn bonds with AFM exchange interaction \cite{bau} which are responsible for the GP formation and are  fixed in the lattice. Thus,  as far as the microscopic origin of disorder-induced interaction and its nature and distribution are concerned, the present system is quite different from those of divalent doped manganites. Furthermore, the model approach shows that the Griffiths phase effect is enhanced in presence of correlated disorder in the JT dominated region and the span of GP region ($T_{GP}$$-$$T_C^R$) shrinks with the increase of hole concentration and disappears at half-doping \cite{bou}.  The increase of ($T_{GP}$$-$$T_C^R$) in La$_{1-x}$Sr$_x$MnO$_3$ with decreasing $x$ and increasing JT distortion is consistent with this picture \cite{deisen}. However, the observed GP characteristics in La$_{0.6}$Sr$_{0.4}$Mn$_{1-x}$Co$_x$O$_3$  are quite different from this prediction because ($T_{GP}$$-$$T_C^R$) for the present system  is maximum ($\sim$155 K for $x$$=$0.40) among the manganites, although, it is close to half-doping and far from the static JT dominated regime. \\

Finally, the appearance of GP in La$_{0.6}$Sr$_{0.4}$CoO$_3$ with Mn doping adds a new dimension to GP phenomenon in DE system. La$_{1-x}$Sr$_x$CoO$_3$ is another DE system, which is similar in many aspects to La$_{1-x}$Sr$_x$MnO$_3$. The nanoscale phase separation and percolative conduction are also quite common in cobaltites. Yet, GP in La$_{1-x}$Sr$_x$CoO$_3$ was not observed earlier. Though, magnetic and neutron diffraction studies show the presence of  FM clusters above $T_C$, they are not Griffiths-like in nature \cite{he}.\\

In conclusion, we  have reported GP in La$_{0.6}$Sr$_{0.4}$Mn$_{1-x}$Co$_{x}$O$_3$ which is markedly different from those observed in other perovskite manganites. GP has been observed in the metallic PM state for $x$$<$0.10 and $x$$>$0.85, refuting the plausible assertion that MI transition occurs concurrently with GP. This is possibly due to the difference in the  nature of disorder-induced exchange interaction and its quenching in the present system from those in  divalent doped manganites.  We have also shown that GP emerges in La$_{0.6}$Sr$_{0.4}$CoO$_3$ when Mn is substituted at Co site.\\

The authors wish to thank D. Vieweg and S. Banerjee for their help for magnetic and energy dispersive x-ray measurements.

\end{document}